\documentclass[twocolumn,showpacs,aps,prl,superscriptaddress]{revtex4}
\usepackage[dvips]{graphicx}
\usepackage{latexsym}
\def\bslantfrac#1#2{{#1}\backslash\kern-0.1em{#2}}

\begin{document}
\newcommand{\Real}{\text{Re}}
\newcommand{\Imag}{\text{Im}}

\title{Scaling Tests of the  Cross Section for Deeply Virtual Compton Scattering}
\author{C.~Mu\~noz~Camacho}
\affiliation{CEA Saclay, DAPNIA/SPhN, F-91191 Gif-sur-Yvette, France}
\author{A.~Camsonne}
\affiliation{Universit\'e Blaise Pascal/CNRS-IN2P3, F-63177 Aubi\`ere, France}
\author{M.~Mazouz}
\affiliation{Laboratoire de Physique Subatomique et de Cosmologie, 38026 Grenoble, France}
\author{C.~Ferdi}
\affiliation{Universit\'e Blaise Pascal/CNRS-IN2P3, F-63177 Aubi\`ere, France}
\author{G.~Gavalian}
\affiliation{Old Dominion University, Norfolk, Virginia 23508, USA}
\author{E.~Kuchina}
\affiliation{Rutgers, The State University of New Jersey, Piscataway, New Jersey 08854, USA}
\author{M.~Amarian}
\affiliation{Old Dominion University, Norfolk, Virginia 23508, USA}
\author{K.~A.~Aniol}
\affiliation{California State University, Los Angeles, Los Angeles, California 90032, USA}
\author{M.~Beaumel}
\affiliation{CEA Saclay, DAPNIA/SPhN, F-91191 Gif-sur-Yvette, France}
\author{H.~Benaoum}
\affiliation{Syracuse University, Syracuse, New York 13244, USA}
\author{P.~Bertin}
\affiliation{Universit\'e Blaise Pascal/CNRS-IN2P3, F-63177 Aubi\`ere, France}
\affiliation{Thomas Jefferson National Accelerator Facility, Newport News, Virginia 23606, USA}
\author{M.~Brossard}
\affiliation{Universit\'e Blaise Pascal/CNRS-IN2P3, F-63177 Aubi\`ere, France}
\author{J.-P.~Chen}
\affiliation{Thomas Jefferson National Accelerator Facility, Newport News, Virginia 23606, USA}
\author{E.~Chudakov}
\affiliation{Thomas Jefferson National Accelerator Facility, Newport News, Virginia 23606, USA}
\author{B.~Craver}
\affiliation{University of Virginia, Charlottesville, Virginia 22904, USA}
\author{F.~Cusanno}
\affiliation{INFN/Sezione Sanit\`{a}, 00161 Roma, Italy}
\author{C.W.~de~Jager}
\affiliation{Thomas Jefferson National Accelerator Facility, Newport News, Virginia 23606, USA}
\author{A.~Deur}
\affiliation{Thomas Jefferson National Accelerator Facility, Newport News, Virginia 23606, USA}
\author{R.~Feuerbach}
\affiliation{Thomas Jefferson National Accelerator Facility, Newport News, Virginia 23606, USA}
\author{J.-M.~Fieschi}
\affiliation{Universit\'e Blaise Pascal/CNRS-IN2P3, F-63177 Aubi\`ere, France}
\author{S.~Frullani}
\affiliation{INFN/Sezione Sanit\`{a}, 00161 Roma, Italy}
\author{M.~Gar\c con}
\affiliation{CEA Saclay, DAPNIA/SPhN, F-91191 Gif-sur-Yvette, France}
\author{F.~Garibaldi}
\affiliation{INFN/Sezione Sanit\`{a}, 00161 Roma, Italy}
\author{O.~Gayou}
\affiliation{Massachusetts Institute of Technology,Cambridge, Massachusetts 02139, USA}
\author{R.~Gilman}
\affiliation{Rutgers, The State University of New Jersey, Piscataway, New Jersey 08854, USA}
\author{J.~Gomez}
\affiliation{Thomas Jefferson National Accelerator Facility, Newport News, Virginia 23606, USA}
\author{P.~Gueye}
\affiliation{Hampton University, Hampton, Virginia 23668, USA}
\author{P.A.M.~Guichon}
\affiliation{CEA Saclay, DAPNIA/SPhN, F-91191 Gif-sur-Yvette, France}
\author{B.~Guillon}
\affiliation{Laboratoire de Physique Subatomique et de Cosmologie, 38026 Grenoble, France}
\author{O.~Hansen}
\affiliation{Thomas Jefferson National Accelerator Facility, Newport News, Virginia 23606, USA}
\author{D.~Hayes}
\affiliation{Old Dominion University, Norfolk, Virginia 23508, USA}
\author{D.~Higinbotham}
\affiliation{Thomas Jefferson National Accelerator Facility, Newport News, Virginia 23606, USA}
\author{T.~Holmstrom}
\affiliation{College of William and Mary, Williamsburg, Virginia 23187, USA}
\author{C.E.~Hyde-Wright}
\affiliation{Old Dominion University, Norfolk, Virginia 23508, USA}
\author{H.~Ibrahim}
\affiliation{Old Dominion University, Norfolk, Virginia 23508, USA}
\author{R.~Igarashi}
\affiliation{University of Saskatchewan, Saskatchewan, Saskatchewan, Canada, S7N 5C6}
\author{X.~Jiang}
\affiliation{Rutgers, The State University of New Jersey, Piscataway, New Jersey 08854, USA}
\author{H.S.~Jo}
\affiliation{Institut de Physique Nucl\'eaire CNRS-IN2P3, Orsay, France}
\author{L.J.~Kaufman}
\affiliation{University of Massachusetts Amherst, Amherst, Massachusetts 01003, USA}
\author{A.~Kelleher}
\affiliation{College of William and Mary, Williamsburg, Virginia 23187, USA}
\author{A.~Kolarkar}
\affiliation{University of Kentucky, Lexington, Kentucky 40506, USA}
\author{G.~Kumbartzki}
\affiliation{Rutgers, The State University of New Jersey, Piscataway, New Jersey 08854, USA}
\author{G.~Laveissi\`ere}
\affiliation{Universit\'e Blaise Pascal/CNRS-IN2P3, F-63177 Aubi\`ere, France}
\author{J.J.~LeRose}
\affiliation{Thomas Jefferson National Accelerator Facility, Newport News, Virginia 23606, USA}
\author{R.~Lindgren}
\affiliation{University of Virginia, Charlottesville, Virginia 22904, USA}
\author{N.~Liyanage}
\affiliation{University of Virginia, Charlottesville, Virginia 22904, USA}
\author{H.-J.~Lu}
\affiliation{Department of Modern Physics, University of Science and Technology of China, Hefei 230026, China}
\author{D.J.~Margaziotis}
\affiliation{California State University, Los Angeles, Los Angeles, California 90032, USA}
\author{Z.-E.~Meziani}
\affiliation{Temple University, Philadelphia, Pennsylvania 19122, USA}
\author{K.~McCormick}
\affiliation{Rutgers, The State University of New Jersey, Piscataway, New Jersey 08854, USA}
\author{R.~Michaels}
\affiliation{Thomas Jefferson National Accelerator Facility, Newport News, Virginia 23606, USA}
\author{B.~Michel}
\affiliation{Universit\'e Blaise Pascal/CNRS-IN2P3, F-63177 Aubi\`ere, France}
\author{B.~Moffit}
\affiliation{College of William and Mary, Williamsburg, Virginia 23187, USA}
\author{P.~Monaghan}
\affiliation{Massachusetts Institute of Technology,Cambridge, Massachusetts 02139, USA}
\author{S.~Nanda}
\affiliation{Thomas Jefferson National Accelerator Facility, Newport News, Virginia 23606, USA}
\author{V.~Nelyubin}
\affiliation{University of Virginia, Charlottesville, Virginia 22904, USA}
\author{M.~Potokar}
\affiliation{Institut Jozef Stefan, University of Ljubljana, Ljubljana, Slovenia}
\author{Y.~Qiang}
\affiliation{Massachusetts Institute of Technology,Cambridge, Massachusetts 02139, USA}
\author{R.D.~Ransome}
\affiliation{Rutgers, The State University of New Jersey, Piscataway, New Jersey 08854, USA}
\author{J.-S.~R\'eal}
\affiliation{Laboratoire de Physique Subatomique et de Cosmologie, 38026 Grenoble, France}
\author{B.~Reitz}
\affiliation{Thomas Jefferson National Accelerator Facility, Newport News, Virginia 23606, USA}
\author{Y.~Roblin}
\affiliation{Thomas Jefferson National Accelerator Facility, Newport News, Virginia 23606, USA}
\author{J.~Roche}
\affiliation{Thomas Jefferson National Accelerator Facility, Newport News, Virginia 23606, USA}
\author{F.~Sabati\'e}
\affiliation{CEA Saclay, DAPNIA/SPhN, F-91191 Gif-sur-Yvette, France}
\author{A.~Saha}
\affiliation{Thomas Jefferson National Accelerator Facility, Newport News, Virginia 23606, USA}
\author{S.~Sirca}
\affiliation{Institut Jozef Stefan, University of Ljubljana, Lujubljana, Slovenia}
\author{K.~Slifer}
\affiliation{University of Virginia, Charlottesville, Virginia 22904, USA}
\author{P.~Solvignon}
\affiliation{Temple University, Philadelphia, Pennsylvania 19122, USA}
\author{R.~Subedi}
\affiliation{Kent State University, Kent, Ohio 44242, USA}
\author{V.~Sulkosky}
\affiliation{College of William and Mary, Williamsburg, Virginia 23187, USA}
\author{P.E.~Ulmer}
\affiliation{Old Dominion University, Norfolk, Virginia 23508, USA}
\author{E.~Voutier}
\affiliation{Laboratoire de Physique Subatomique et de Cosmologie, 38026 Grenoble, France}
\author{K.~Wang}
\affiliation{University of Virginia, Charlottesville, Virginia 22904, USA}
\author{L.B.~Weinstein}
\affiliation{Old Dominion University, Norfolk, Virginia 23508, USA}
\author{B.~Wojtsekhowski}
\affiliation{Thomas Jefferson National Accelerator Facility, Newport News, Virginia 23606, USA}
\author{X.~Zheng}
\affiliation{Argonne National Laboratory, Argonne, Illinois, 60439, USA}
\author{L.~Zhu}
\affiliation{University of Illinois, Urbana, Illinois 61801, USA}
\collaboration{The Jefferson Lab Hall A Collaboration}

\begin{abstract}
We present the first measurements of the $\vec{e}p\rightarrow ep\gamma$ cross 
section in the deeply virtual Compton scattering (DVCS) regime and the 
valence quark region. The $Q^2$ dependence (from 1.5 to 2.3 GeV$^2$) of 
the helicity-dependent cross section indicates the twist-2 dominance of 
DVCS, proving that generalized parton distributions (GPDs) are accessible 
to experiment at moderate $Q^2$. The helicity-independent cross section is 
also measured at $Q^2=2.3\,$GeV$^2$. We present the first 
model-independent measurement of linear combinations of GPDs and GPD 
integrals up to the twist-3 approximation.
\end{abstract}


\maketitle

\begin{figure}
\begin{center}
\includegraphics[width=\linewidth]{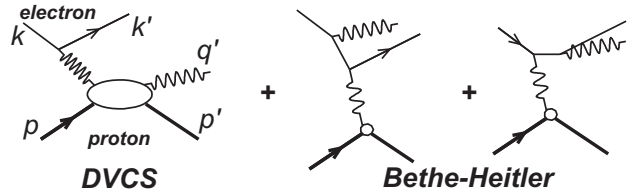}
\caption{ 
Lowest-order QED diagrams for the process $ep\rightarrow ep\gamma$,
including the DVCS and Bethe-Heitler (BH) amplitudes. The external
momentum four-vectors are defined on the diagram. The virtual photon
momenta are $q=k-k'$ in the DVCS- and $\Delta=q-q'$ in the BH-amplitudes.
The invariants are: $W^2=(q+p)^2$, $Q^2=-q^2>0$, $t=\Delta^2$, $x_{\rm
Bj}=Q^2/(2 p\cdot q)$, and the DVCS scaling variable
$\xi=-\overline{q}^2/(\overline{q}\cdot P)\approx x_{\rm Bj}/(2-x_{\rm
Bj})$, with $\overline{q}= (q+q')/2$ and $P=p+p'$.
}
\label{fig:1}
\end{center}
\end{figure}

Measurements of electro-weak form factors determine nucleon spatial 
structure, and deep inelastic scattering (DIS) of leptons off the nucleon 
measures parton distribution functions, which determine longitudinal 
momentum distributions. The demonstration by Ji~\cite{Ji:1996ek}, 
Radyushkin~\cite{Radyushkin:1997ki}, and Mueller {\it et 
al.\/}~\cite{Mueller:1998fv}, of a formalism to relate the spatial and 
momentum distributions of the partons allows the exciting possibility of 
determining spatial distributions of quarks and gluons in the nucleon as a 
function of the parton wavelength. These new structure functions, now 
called generalized parton distributions (GPD), became of experimental 
interest when it was shown~\cite{Ji:1996ek}{} that they are accessible 
through deeply virtual Compton scattering (DVCS) and its interference with 
the Bethe-Heitler (BH) process (Fig.~\ref{fig:1}). Figure~\ref{fig:1} 
presents our kinematic nomenclature. DVCS is defined kinematically by the 
limit $-t\ll Q^2$ and $Q^2$ much larger than the quark confinement scale.

\begin{table*}[floatfix]
\caption{
Experimental $ep\rightarrow e p \gamma$ kinematics, for incident beam 
energy $E=5.75$ GeV. $\theta_q$ is the central value of the {\bf q}-vector 
direction. The PbF$_2$ calorimeter was centered on $\theta_q$ for each 
setting. The photon energy for ${\bf q'}\parallel{\bf q}$ is $E_\gamma$.
}
\begin{ruledtabular}
\begin{tabular}{cccccccc}
Kin & $k'$ (GeV/c) & $\theta_e$ $({}^\circ)$ & $Q^2$ (GeV$^2$) & $x_{\rm
  Bj}$ & $\theta_q$ $({}^\circ)$ & $W$ (GeV) & $E_\gamma$ (GeV)\\
\hline
1 & 3.53 & 15.6 & { 1.5} & 0.36 & $ -22.3$ & 1.9 & 2.14\\
2 & 2.94  & 19.3 & { 1.9} & 0.36 & {$ -18.3$}& 2.0 & 2.73\\
3 & 2.34 & 23.8 & { 2.3} & 0.36 & {$ -14.8$}& 2.2 & 3.33\\
\end{tabular}
\end{ruledtabular}
\label{tab:DVCSkin}
\end{table*}

The factorization proofs~\cite{Collins:1998be,Ji:1998xh} confirmed the 
connection between DVCS and DIS. Diehl {\it et al.\/}~\cite{Diehl:1997bu} 
showed that the twist-2 and twist-3 contributions in the DVCS-BH 
interference terms (the first two leading orders in $1/Q$) could be 
extracted independently from the azimuthal-dependence of the 
helicity-dependent cross section. Burkardt~\cite{Burkardt:2000za} showed 
that the $t$-dependence of the GPDs is the Fourier conjugate to the 
transverse spatial distribution of quarks in the infinite momentum frame 
as a function of momentum fraction. Ralston and 
Pire~\cite{Ralston:2001xs}, Diehl~\cite{Diehl:2002he} and Belitsky {\it et 
al.\/}~\cite{Belitsky:2003nz} extended this interpretation to the general 
case of skewness $\xi\ne0$. The light-cone wave function representation by 
Brodsky {\it et al}~\cite{Brodsky:2000xy} allows GPDs to be interpreted as 
interference terms of wave functions for different parton configurations 
in a hadron.

These concepts stimulated an intense experimental effort in DVCS. The 
H1~\cite{Aktas:2005ty,Adloff:2001cn} and ZEUS~\cite{Chekanov:2003ya} 
collaborations measured the cross section for $x_{\rm Bj}\approx 10^{-3}$. 
The HERMES collaboration measured relative 
beam helicity~\cite{Airapetian:2001yk} and beam-charge 
asymmetries~\cite{Ellinghaus:2002bq,unknown:2006zr}. Relative 
beam-helicity~\cite{Stepanyan:2001sm} and longitudinal 
target~\cite{Chen:2006na} asymmetries were measured at the Thomas 
Jefferson National Accelerator Facility (JLab) by the CLAS collaboration.

Extracting GPDs from DVCS requires the fundamental demonstration that DVCS 
is well described by the twist-2 diagram of Fig.~\ref{fig:1} at finite 
$Q^2$. This letter reports the first strong evidence of this cornerstone 
hypothesis, necessary to validate all previous and future GPD measurements 
using DVCS. We present the determination of the cross section of the 
$\vec{e}p\rightarrow ep\gamma$ reaction for positive and negative electron 
helicity in the kinematics of Table~\ref{tab:DVCSkin}.

The E00-110~\cite{e00-110} experiment ran in Hall A~\cite{Alcorn:2004sb} 
at JLab. The 5.75 GeV electron beam was incident on a 15 cm liquid H$_2$ 
target. Our typical luminosity was $10^{37}$/cm$^2$/s with $76\%$ beam 
polarization. We detected scattered electrons in one high resolution 
spectrometer (HRS).  Photons above a 1~GeV energy threshold (and 
$\gamma\gamma$ coincidences from $\pi^0$ decay) were detected in a 
$11\times 12$ array of $3 \times 3 \times 18.6$ cm$^3$ PbF$_2$ crystals, 
whose front face was located 110 cm from the target center. We calibrated 
the PbF$_2$ array by coincident elastic H$(e,e_{\rm Calo}^\prime p_{\rm 
HRS})$ data. With (elastic) $k'=4.2$~GeV/c, we obtain a PbF$_2$ resolution 
of 2.4\% in energy and 2~mm in transverse position (one-$\sigma$). The 
calibration was monitored by reconstruction of the $\pi^0\rightarrow 
\gamma\gamma$ mass from H$(e,e'\pi^0)X$ events.

\begin{figure}
\begin{center}
\includegraphics[width=\linewidth]{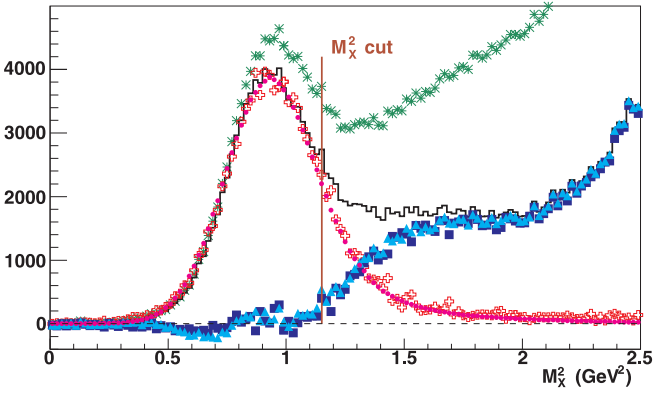}
\caption{
(color online). Missing mass squared for H$(e,e'\gamma)X$ events (stars) at $Q^2=2.3$ 
GeV$^2$ and $-t\in[0.12,0.4]$ GeV$^2$, integrated over the azimuthal angle 
of the photon $\phi_{\gamma\gamma}$. The solid histogram shows the data 
once the H$(e,e'\gamma)\gamma X'$ events have been subtracted. The other 
histograms are described in the text.
}
\label{fig:mm2}
\end{center}
\end{figure}

We present in Fig.~\ref{fig:mm2} the missing mass squared obtained for 
H$(e,e'\gamma)X$ events, with coincident electron-photon detection. After 
subtraction of an accidental coincidence sample, we have the following 
competing channels in addition to
H$(e,e'\gamma) p$\,:  $e p \rightarrow e\pi^0 p$,
$e p \rightarrow e \pi^0 N\pi$,
$e p \rightarrow e \gamma N\pi$,
$e p \rightarrow e \gamma N\pi\pi\ldots$.
From symmetric (lab-frame) $\pi^0$-decay, we obtain a high statistics 
sample of H$(e,e'\pi^0)X'$ events, with two photon clusters in the PbF$_2$ 
calorimeter. From these events, we determine the statistical sample of 
[asymmetric] H$(e,e'\gamma)\gamma X'$ events that must be present in our 
H$(e,e'\gamma)X$ data. The solid $M_X^2$ spectrum displayed in 
Fig.~\ref{fig:mm2} was obtained after subtracting this $\pi^0$ yield from 
the total (stars) distribution. This is a $14\%$ average subtraction in 
the exclusive window defined by $M_X^2$ cut in Fig.~\ref{fig:mm2}. 
Depending on the bin in $\phi_{\gamma\gamma}$ and $t$, this subtraction 
varies from 6\% to 29\%. After our $\pi^0$ subtraction, the only remaining 
channels, of type H$(e,e'\gamma)N\pi$, $N\pi\pi$, {\it etc.\/} are 
kinematically constrained to $M_X^2 > (M+m_\pi)^2$. This is the value 
($M_X^2$ cut in Fig.~\ref{fig:mm2}) we chose for truncating our 
integration. Resolution effects can cause the inclusive channels to 
contribute below this cut. To evaluate this possible contamination, we 
used an additional proton array (PA) of 100 plastic scintillators. The PA 
subtended a solid angle (relative to the nominal direction of the {\bf 
q} vector) of $18^\circ<\theta_{\gamma p}<38^\circ$ and $45^\circ < 
\phi_{\gamma p} = 180^\circ-\phi_{\gamma\gamma} < 315^\circ$, arranged in 
5 rings of 20 detectors. For H$(e,e'\gamma)X$ events near the exclusive 
region, we can predict which block in the PA should have a signal from a 
proton from an exclusive H$(e,e'\gamma p)$ event. Open crosses show the 
$X=(p+y)$ missing mass squared distribution for H$(e,e'\gamma p)y$ events 
in the predicted PA block, with a signal above an effective threshold $30$ 
MeV. Squares show our inclusive yield, obtained by subtracting the 
normalized triple coincidence yield from the H$(e,e'\gamma)X$ yield. The 
dotted curve shows our simulated H$(e,e'\gamma)p$ spectrum, including 
radiative and resolution effects, normalized to fit the data for $M_X^2\le 
M^2$. Triangles show the estimated inclusive yield obtained by subtracting 
the simulation from the data. Squares and triangles are in good agreement, 
and show that our exclusive yield has less than $3\%$ contamination from 
inclusive processes.

To order twist-3 the DVCS helicity-dependent ($d\Sigma$) and 
helicity-independent ($d\sigma$) cross sections 
are~\cite{Belitsky:2001ns}:
\begin{eqnarray}
{d^4\Sigma \over d^4\Phi}  
&=& {1\over 2}
  \left[ {d^4\sigma^+\over d^4\Phi} - {d^4\sigma^-\over d^4\Phi}\right] 
     = {d^4\Sigma(|DVCS|^2)\over d^4\Phi} \nonumber \\
&+&  \sin(\phi_{\gamma\gamma})
  \Gamma_{1}^{\text{Im}} \,
  \text{Im} \left[{\mathcal C^I}({\mathcal F})\right]  \nonumber \\
 &-&\sin(2\phi_{\gamma\gamma}) 
     \Gamma_{2}^{\text{Im}} \,
    {\text{Im}}\left[{\mathcal C^I}({\mathcal F}^{\rm eff})\right]\,,  
\label{eq:dSigma}\\
{d^4\sigma\over d^4\Phi}    
&=& {1\over 2}
     \left[{d^4\sigma^+\over d^4\Phi} + {d^4\sigma^-\over d^4\Phi}\right]
        =  {d^4\sigma(|DVCS|^2) \over d^4\Phi}  \nonumber \\
&+& {d^4\sigma(|BH|^2) \over d^4\Phi}
  +  \Gamma_{0,\Delta}^{\Real}  {\Real \text{e}}\left[
              {\mathcal C^I}+\Delta{\mathcal C^I}\right] ({\mathcal F}) 
    \nonumber \\
  &+&  \Gamma_0^{\Real}  {\Real \text{e}}\left[{\mathcal C^I}({\mathcal F})\right] 
-\cos(\phi_{\gamma\gamma})
  \Gamma_{1}^{\Real} {\Real}\left[{\mathcal C^I}({\mathcal F})\right]
   \nonumber \\
 &+& 
   \cos(2\phi_{\gamma\gamma}) 
     \Gamma_{2}^{\Real}
    {\Real} \left[{\mathcal C^I}({\mathcal F}^{\rm eff})\right]\,,
\label{eq:dsigma}
\end{eqnarray}
where $d^4\Phi=dQ^2 dx_{\rm Bj} dt d\phi_{\gamma\gamma}$ and the azimutal 
angle $\phi_{\gamma\gamma}$ of the detected photon follows the 
``Trento-Convention''~\cite{Bacchetta:2004jz}. The $\Gamma_{n}^{\Real,\Imag}$ 
are kinematic factors with a $\phi_{\gamma\gamma}$ dependence that arises 
from the electron propagators of the BH amplitude. The ${\mathcal C^I}$ 
and $\Delta{\mathcal C^I}$ angular harmonics depend on the interference of 
the BH amplitude with the set
${\mathcal F}=
\{{\mathcal H,\,\mathcal E,\,\tilde{\mathcal H},\,\tilde{\mathcal E}}\}$
of twist-2 Compton form factors (CFFs) or the related set
${\mathcal F}^{\rm eff}$ of effective twist-3 CFFs:
\begin{eqnarray}
{\mathcal C^I}({\mathcal F}) = F_1{\mathcal H}+
     \xi G_M \tilde{\mathcal H}
 -{t\over 4 M^2} F_2{\mathcal E}\quad\quad\ \ \,\,\,\ \ 
\label{eq:gpds1} \\
{\mathcal C^I}({\mathcal F}^{\rm eff}) 
  = F_1{\mathcal H}^{\rm eff}+
      \xi G_M\tilde{\mathcal H}^{\rm eff}
   -{t\over 4 M^2} F_2 {\mathcal E}^{\rm eff}\ \ \   
\label{eq:gpds2}\\
\left[{\mathcal C^I}+\Delta{\mathcal C^I}\right]({\mathcal F}) =
           F_1{\mathcal H}-{t\over 4 M^2} F_2{\mathcal E} 
   - \xi^2 G_M 
       \left[{\mathcal H}+{\mathcal E}\right].\  
\label{eq:gpds3}
\end{eqnarray}
$F_1(t)$, $F_2(t)$ and
$G_M(t)=F_1(t)+F_2(t)$ are the elastic form factors. 
CFFs are defined in terms of the GPDs $H_f$, $E_f$,
$\tilde{H}_f$, and $\tilde{E}_f$, defined for each quark flavor $f$. 
For example ($f\in\{u,d,s\}$):
\begin{eqnarray}
{\mathcal H}(\xi,t) &=& \sum_{f} \left[\frac{e_f}{e}\right]^2 \Biggl\{
         i\pi     \left[H_f(\xi,\xi,t) - H_f(-\xi,\xi,t)\right]
 \nonumber \\ 
  &+& 
 {\mathcal P} 
    \int_{-1}^{+1} dx
\left[ {2x \over \xi^2-x^2 } \right] 
H_f(x,\xi,t)\Biggr\}.
\label{eq:CFF}
\end{eqnarray}

Thus, the DVCS helicity-dependent and helicity-independent cross sections 
provide very distinct and complementary information on GPDs. On one hand, 
$d\Sigma$ measures the imaginary part of the BH-DVCS interference terms 
and provides direct access to GPDs at $x=\xi$. On the other hand, 
$d\sigma$ determines the real part of the BH-DVCS interference terms and 
measures the integral of GPDs over their full domain in $x$. This real 
part of the BH-DVCS interference term is the same interference term that 
can be obtained by measurements of the difference of electron and positron 
(or $\mu^\pm$) DVCS cross sections.

The twist-2 and twist-3 CFFs are matrix elements of quark-gluon operators 
and are independent of $Q^2$ (up to logarithmic QCD evolution). Their 
$Q^2$ variation measures the potential contamination from higher twists.

\begin{table*}[t]
\caption{Angular Harmonics fit results, $\Imag$ and $\Real$ parts, and
  their statistical uncertainties.}
\begin{ruledtabular}
\begin{tabular}{ccrrrr}
\multicolumn{1}{c}{
$\displaystyle\bslantfrac{ Q^2}{\langle t \rangle}$
(GeV$^2$)} & &$t=-0.33$ &$-0.28$& $-0.23$&$-0.17$\\
\hline
$\Imag$ &1.5
& $2.1\pm0.3$ & $2.1\pm0.3$ & $2.0\pm0.2$ & $3.2\pm0.2$ \\
$[{\mathcal C}^{\mathcal I}(\mathcal F)]$ &1.9
& $1.9\pm0.2$ & $2.3\pm0.2$ & $2.5\pm0.2$ & $3.2\pm0.2$ \\
$~$&2.3
& $2.1\pm0.2$ & $2.4\pm0.2$ & $2.6\pm0.2$ & $3.3\pm0.3$ \\
$\Imag$&1.5
& $2.8\pm2.0$ & $2.5\pm2.0$ & $0.1\pm2.1$ & $0.6\pm2.4$ \\
$[{\mathcal C}^{\mathcal I}(\mathcal F^{\text{eff}})]$&1.9
& $0.3\pm1.4$ & $3.8\pm1.5$ &$-0.9\pm1.8$ & $4.7\pm2.7$ \\
$~$&2.3
& $5.3\pm1.6$ & $0.7\pm1.8$ & $0.2\pm2.5$ & $4.0\pm4.6$ \\
& \multicolumn{5}{c}{$\qquad Q^2=2.3$ GeV$^2$, $\Real$ part of Angular Harmonics}
\\ 
\multicolumn{1}{c}{${\mathcal C}({\mathcal F})$} &  & $-2.4\pm0.1$
                 & $-2.0\pm0.1$ & $-1.7\pm0.1$ & $-0.7\pm0.2$ \\
\multicolumn{1}{c}
    {$\left[{\mathcal C}+\Delta{\mathcal C}\right]({\mathcal F})$}
     &  & $0.1\pm0.1$
                 & $ 0.8\pm0.1$ & $ 1.6\pm0.1$ & $ 2.5\pm0.1$ \\
\multicolumn{1}{c}{$[{\mathcal C}({\mathcal F}^{\text{eff}})] $} & & 
    $-1.4\pm0.5$ 
                 & $ 0.6\pm0.6$ & $ 1.0\pm0.8$ & $3.4\pm1.4$ \\
\end{tabular}
\end{ruledtabular}
\label{tab:AmpDiff}
\end{table*}

The helicity-independent cross section also has a 
$\cos(3\phi_{\gamma\gamma})$ twist-2 gluon transversity term. We expect 
this term to be small, and do not include it in our analysis. We neglect 
the DVCS$^2$ terms in our analysis. Therefore, our results for 
$\Imag[{\mathcal C}^{\mathcal I}]$ and $\Real[{\mathcal C}^{\mathcal I}]$ 
may contain, respectively, twist-3 and twist-2 DVCS$^2$ terms, which enter 
with similar $\phi_{\gamma\gamma}$ dependence. However, the DVCS$^2$ terms 
in both $d\sigma$ and $d\Sigma$ are kinematically suppressed by at least 
an order of magnitude in our kinematics~\cite{Belitsky:2001ns}, because 
they are not enhanced by the BH amplitude. In any case, the terms we 
neglect do not affect the cross sections we extract, which are accurately 
parametrized, within statistics, by the contributions we included.

Our simulation includes internal bremsstrahlung in the scattering process 
and external bremsstrahlung and ionization straggling in the target and 
scattering chamber windows. We include spectrometer resolution and 
acceptance effects and a full GEANT3 simulation of the detector response 
to the DVCS photons and protons. The spectrometer acceptance is defined 
for both the data and simulation by a $R$-function 
cut~\cite{Rvachev:2001}. Radiative corrections for virtual photons and 
unresolved real photons are applied according to the VCS (BH+Born 
amplitude) specific prescriptions of Ref.~\cite{Vanderhaeghen:2000ws}. 
This results in a global correction factor (independent of 
$\phi_{\gamma\gamma}$ or helicity) of $0.91\pm0.02$ applied to our 
experimental yields. Within the quoted uncertainty, this correction is 
independent of the kinematic setting.

For each $(Q^2, x_{\rm Bj}, t)$ bin, we fit
the $\Real$ and $\Imag$ parts (as appropriate) of the harmonics
${\mathcal C}_n \in \{
{\mathcal C^{\mathcal I}(\mathcal F)},
{\mathcal C^{\mathcal I}(\mathcal F^{\rm eff})},
\left[{\mathcal C^{\mathcal I}}+\Delta{\mathcal C^{\mathcal I}}\right]
({\mathcal F})\}$ as independent parameters. 
In Kin-1 and Kin-2, due to the lower photon energy $E_\gamma$ 
(Table~\ref{tab:DVCSkin}), our acceptance, trigger, and readout did not 
record a comprehensive set of $e p \rightarrow e \pi^0X$ events.  For 
those events we were able to reconstruct, we found only a few percent 
contribution to $d\Sigma$, but a larger contribution to $d\sigma$.  For 
Kin-1,2, we only present results on $d\Sigma$. Our systematic errors in 
the cross-section measurements are dominated by the following 
contributions: $3\%$ from HRS$\times$PbF$_2$ acceptance and luminosity; 
$3\%$ from H$(e,e'\gamma)\gamma X$ ($\pi^0$) background; $2\%$ from 
radiative corrections; and $3\%$ from inclusive H$(e,e'\gamma)N\pi\ldots$ 
background.  The total, added in quadrature, is $5.6\%$. The $d\Sigma$ 
results contain an additional $2\%$ systematic uncertainty from the beam 
polarization. In order to compute the BH contribution in the $d\sigma$ 
analysis we used Kelly's parametrization of form 
factors~\cite{Kelly:2004hm}, which reproduce elastic cross-section world 
data in our $t$ range with 1\% error and 90\% CL.

\begin{figure}[b]
\includegraphics[width=\linewidth]{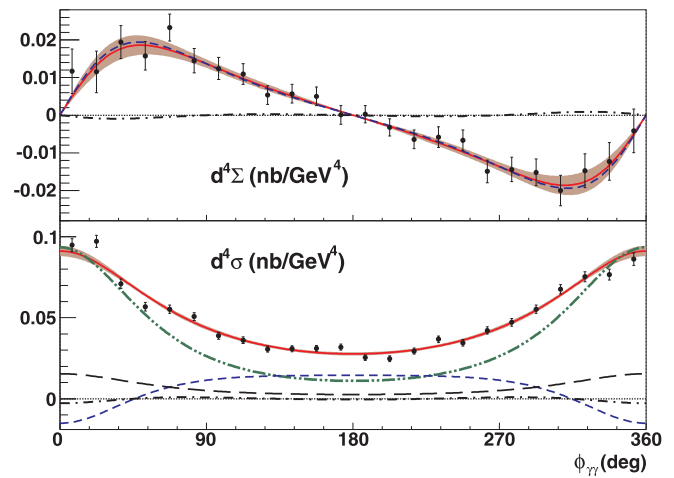}
\caption{
(color online). Data and fit to $d^4\Sigma/[dQ^2 dx_{\rm Bj} dt d\phi_{\gamma\gamma}]$, and 
$d^4\sigma/[dQ^2 dx_{\rm Bj} dt d\phi_{\gamma\gamma}]$, as a function of 
$\phi_{\gamma\gamma}$. Both are in the bin $\langle Q^2, 
t\rangle=(2.3,-0.28)$ GeV$^2$ at $\left\langle x_{\rm Bj} 
\right\rangle=0.36$. Error bars show statistical uncertainties. Solid 
lines show total fits with one-$\sigma$ statistical error bands. 
Systematic uncertainty is given in the text. The dot-dot-dashed line is 
the $|{\rm BH}|^2$ contribution to $d^4\sigma$. The short-dashed lines in 
$d^4\Sigma$ and $d^4\sigma$ are the fitted $\Imag$ and $\Real$ parts of 
${\mathcal C}^{\mathcal I}({\mathcal F})$, respectively. The long-dashed 
line is the fitted $\Real[{\mathcal C}^{\mathcal I}+\Delta\mathcal 
C^{\mathcal I}](\mathcal F)$ term.  The dot-dashed curves are the fitted 
$\Imag$ and $\Real$ parts of ${\mathcal C}^{\mathcal I}({\mathcal F}^{\rm 
eff})$.
}
\label{fig:SigmaDiff}
\end{figure}

\begin{figure}
\includegraphics[width=\linewidth]{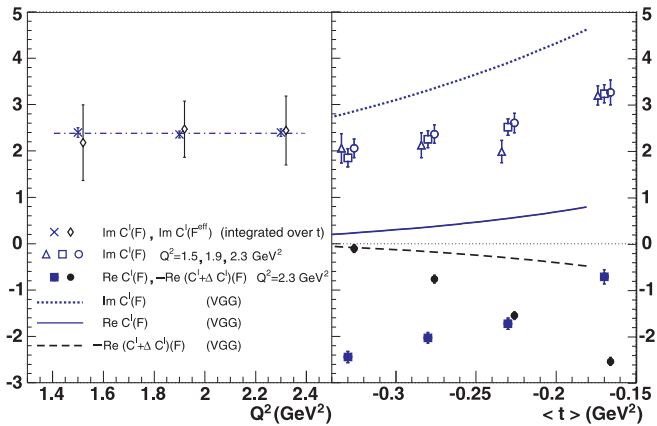}
\caption{
(color online) Left:  $Q^2$ dependence of $\Imag$ parts of (twist-2) ${\mathcal
C}^{\mathcal I}(\mathcal F)$ and (twist-3) ${\mathcal C}^{\mathcal
I}(\mathcal F^{\rm eff})$ angular harmonics, averaged over $t$. The
horizontal line is the fitted average of $\Imag[{\mathcal C}^{\mathcal
I}(\mathcal F)]$. Right : Extracted real and imaginary parts of the twist-2 angular
harmonics as functions of $t$.  The VGG model curves are described in the
text. Note the sign of $-[\mathcal C^{\mathcal I}+\Delta{\mathcal
C}^{\mathcal I}](\mathcal F)$ (data and VGG). Superposed points in both
panels are offset for visual clarity. Their error bars show statistical
uncertainties.
}
\label{fig:coef}
\end{figure}

For one $(Q^2,x_{\rm Bj}, t)$ bin, Fig.~\ref{fig:SigmaDiff} shows the 
helicity-dependent and helicity-independent cross sections, respectively. 
We notice that the twist-3 terms make only a very small contribution to 
the cross sections. Note also that $d\sigma$ is much larger than the BH 
contribution alone, especially from 90$^\circ$ to 270$^\circ$. This 
indicates that the relative Beam Spin Asymmetry $BSA = 
d^4\Sigma/d^4\sigma$ cannot be simply equated to the imaginary part of the 
BH-DVCS interference divided by the BH cross section. 
Table~\ref{tab:AmpDiff} lists the extracted angular harmonics. 
Figure~\ref{fig:coef} (Left) shows the $Q^2$ dependence of the imaginary 
angular harmonic $\Imag[{\mathcal C}^{\mathcal I}]$ over our full $t$ 
domain, with $\langle t\rangle=-0.25$~GeV$^2$ ($\langle t\rangle$ varying 
by $\pm 0.01$~GeV$^2$ over Kin 1--3).

The absence of $Q^2$ dependence of $\Imag[{\mathcal C}^{\mathcal 
I}(\mathcal F)]$ within its 3\% statistical uncertainty provides crucial 
support for the dominance of twist-2 in the DVCS amplitude. Indeed, it 
sets an upper limit $\le 10\%$ to twist-4 and higher contributions. 
$\Imag[{\mathcal C}^{\mathcal I}(\mathcal F)]$ is thereby a direct 
measurement of a linear combination of GPDs. The two twist-2 angular 
harmonics extracted from $d\sigma$ determine distinct combinations of GPD 
integrals, providing most valuable complementary information on GPDs. As 
noted above, the angular harmonic terms in Table~\ref{tab:AmpDiff} may 
include contributions from kinematically suppressed bilinear DVCS$^2$ 
terms omitted in our analysis. In our experiment the acceptance-averaged 
ratios of the kinematic coefficients of the DVCS$^2$ terms to the BH-DVCS 
terms are below 1.2\% for $d\Sigma$ and below 4.5\% for $d\sigma$. The 
cross-section measurements we present are accurate, to the quoted 
uncertainty, and not sensitive within statistics to the neglected terms in 
their harmonic analysis.

Figure~\ref{fig:coef} (Right) displays the twist-2 angular harmonics of 
Table~\ref{tab:AmpDiff} ($\Real$ and $\Imag$ parts) as functions of $t$, 
together with the predictions from a model by Vanderhaeghen, Guichon and 
Guidal (VGG)~\cite{Vanderhaeghen:1999xj,Goeke:2001tz,Guidal:2004nd}. The 
VGG model (twist-2 contribution only, profile parameter 
$b_{val}=b_{sea}=1$, Regge parameter $\alpha'=0.8$ GeV$^{-2}$, GPD 
$E_f=0$) is in qualitative agreement with the $\Imag[\mathcal C^{\mathcal 
I}(\mathcal F)]$ data, but significantly under-predicts the 
principal-value integrals ($\Real$ parts of the angular harmonics).

In summary, we present the first explicit demonstration of exclusivity in 
DVCS kinematics. We also present the first measurements of DVCS cross 
section in the valence quark region. From the $Q^2$ dependence of the 
angular harmonics of the helicity-dependent cross section, we provide 
solid evidence of twist-2 dominance in DVCS, which makes GPDs accessible 
to experiment even at modest $Q^2$. This result supports the striking 
prediction of perturbative QCD scaling in 
DVCS~\cite{Ji:1996ek,Radyushkin:1997ki}. As a consequence of this evidence 
for scaling in the exclusive channel, and our separate determination of 
the helicity-dependent and helicity-independent cross sections, we extract 
for the first time a model-independent combination of GPDs and GPDs 
integrals.

\begin{acknowledgments}
We acknowledge essential work of the JLab accelerator staff and the Hall A 
technical staff. This work was supported by DOE contract 
DOE-AC05-06OR23177 under which the Jefferson Science Associates, LLC, 
operates the Thomas Jefferson National Accelerator Facility. We 
acknowledge additional grants from the DOE and NSF and the CNRS and 
Commissariat \`a l'Energie Atomique.
\end{acknowledgments}

\bibliography{Compton}

\end{document}